\documentclass[a4paper, fleqn]{cas-dc}

\usepackage{upgreek} 
\usepackage[numbers]{natbib}
\usepackage{ulem}
\usepackage{xcolor}
\usepackage{comment}
\usepackage{booktabs}       
\usepackage{siunitx}        
\usepackage{threeparttable} 
\usepackage{multirow}       
\usepackage{tabularx}       

\usepackage{array}
\newcolumntype{C}{>{\centering}p{4em}} 

\def\tsc#1{\csdef{#1}{\textsc{\lowercase{#1}}\xspace}}
\tsc{WGM}
\tsc{QE}
\definecolor{royalpurple}{rgb}{0.47, 0.32, 0.66}

\begin{document}
    \let\WriteBookmarks\relax \def\floatpagepagefraction{1} \def\textpagefraction{.001}

    \shorttitle{Development and Evaluation of a CNN-Based Charged-Particle Event Rejection Algorithm for Soft X-ray Detection in a pnCCD-Based Satellite System}

    \shortauthors{Shen et al.}

    \title[mode = title]{Development and Evaluation of a CNN-Based Charged-Particle Event Rejection Algorithm for Soft X-ray Detection in a pnCCD-Based Satellite System}



    %

    %
    \author[1]{Hsien-Chieh Shen}


    \fnmark[*]

    \ead{hsien-chieh.shen@riken.jp}



    \author[2]{Ryuji Kondo}

    \fnmark[*]

    \ead{ryuji_kondo07@stu.kanazawa-u.ac.jp}



    \cortext[1]{Corresponding author} \makeatletter\def\Hy@Warning#1{}\makeatother

    \author[2]{Makoto Arimoto}
    \author[2]{Tatsuro Kanenaga}
    \author[3]{Hiro Otsuka}
    \author[2]{Shutaro Ueda}
    \author[4]{Junko Hiraga}
    \author[2]{Daisuke Yonetoku}
    \author[2]{Tatsuya Sawano}
    \author[3]{Takanori Sakamoto}
    \author[5]{Hiroshi Tomida}
    \author[5]{Akihiro Doi}
    \author[6]{Hiroshi Nakajima}
    \author[7]{Takaaki Tanaka}
    \author[8]{Robert Hartmann}
    \author[8]{Lothar Strüder}

    \affiliation[1]{organization={RIKEN}, addressline={2-1 Hirosawa}, city={Wako},
    postcode={351-0198}, state={Saitama}, country={Japan}}
    
    \affiliation[2]{organization={Kanazawa University}, addressline={Kakuma-machi}, city ={Kanazawa},
    postcode={920-1192}, state={Ishikawa}, country={Japan}}

    \affiliation[3]{organization={Aoyama Gakuin University}, addressline ={5-10-1, Fuchinobe, Chuo-ku}, City ={Sagamihara}.,
    postcode={252-5258}, state={Kanagawa}, country={Japan}}

    \affiliation[4]{organization={Kwansei Gakuin University}, addressline ={1 Gakuen Uegahara}, City ={Sanda}.,
    postcode={669-1330}, state={Hyogo}, country={Japan}}

    \affiliation[5]{organization={Institute of Space and Astronautical Science}, addressline ={3-1-1 Yoshinodai, Chuo-ku}, City ={Sagamihara}.,
    postcode={252-5210}, state={Kanagawa}, country={Japan}}
    
    \affiliation[6]{organization={Kanto Gakuin University}, addressline ={1-50-1 Mutsuura-higashi, Kanazawa-ku}, City ={Yokohama}.,
    postcode={236-8501}, state={Kanagawa}, country={Japan}}

    \affiliation[7]{organization={Konan University}, addressline ={8-9-1 Okamoto, Higashinada-ku}, City ={Kobe}.,
    postcode={658-8501}, state={Hyogo}, country={Japan}}

    \affiliation[8]{organization={PNSensor GmbH.}, addressline ={Otto-Hahn-Ring 6 }, City ={München}.,
    postcode={81739}, state={Bavaria}, country={Germany}}


    \begin{abstract}
All-sky surveys in the soft X-ray band are essential for detecting transient objects such as high-redshift gamma-ray bursts (GRBs), which provide key insights into the early universe. 
HiZ-GUNDAM is a future satellite mission designed to explore the early universe by detecting and localizing high-redshift GRBs. 
Its wide-field X-ray monitor, EAGLE, combines Lobster Eye Optics with a pnCCD imaging detector operating in the 0.4--4~keV band.
Because of limited telemetry in satellite operations, full-frame pnCCD image data cannot be downlinked, requiring onboard event selection and data reduction.
However, charged particles in the space environment produce background events that can be misidentified as X-ray photons, degrading detection sensitivity and potentially triggering false alerts. 
In this study, we developed a pnCCD readout system and evaluated charged-particle rejection algorithms, including conventional grade methods and a convolutional neural network (CNN)-based approach. 
Performance was evaluated using X-ray events from an $^{55}$Fe source and electron events from a $^{90}$Sr $\beta$ source.
The results show that the CNN method significantly improves discrimination performance, reducing the misclassification rate from 10.2--11.9\% for conventional grade methods to 3.1\%, while maintaining a high acceptance rate for X-ray events.
The improvement is particularly pronounced at higher deposited energies, where charged-particle events are more likely to be misidentified by threshold-based methods. 
This reflects the ability of the CNN approach to capture detailed spatial features of charge distributions, distinguishing track-like particle events from X-ray events.
A comparison with a thin-depletion-layer CMOS sensor further indicates that the thicker depletion layer of the pnCCD enhances discrimination performance.
These results demonstrate that CNN-based event classification can substantially reduce charged-particle contamination while maintaining high X-ray acceptance, making it a promising approach for onboard event selection in future pnCCD-based wide-field X-ray missions.


    \end{abstract}




    \maketitle

\section{Introduction}
\label{sec:Intro}
    Gamma-ray bursts (GRBs) are the most energetic explosive phenomena in the universe, releasing enormous energies of $10^{52}$--$10^{54}$ erg within durations ranging from tens of milliseconds to several hundred seconds. 
    Many GRBs are observed at cosmological distances, including events originating from the early universe. 
    These high-redshift GRBs are expected to be among the brightest X-ray sources illuminating the distant universe and can serve as powerful probes of the early universe.
    \textit{HiZ-GUNDAM} \cite{DY2025HiZGUNDAM} is a future satellite mission planned for launch in the 2030s, aiming to explore the early universe by utilizing GRBs as cosmological probes. 
    Since GRBs occur unpredictably in time and location, a wide-field instrument capable of continuously monitoring a large portion of the sky is essential. 
    In addition, high-redshift GRBs are observed preferentially in the soft X-ray band due to cosmological redshift effects. 
    Based on these scientific requirements, the wide-field X-ray monitor EAGLE \cite{MA2026} onboard \textit{HiZ-GUNDAM} employs Lobster Eye Optics \cite{Angel1979} combined with a pixel focal-plane detector. 
    EAGLE consists of 16 modules, each covering approximately $11^\circ \times 11^\circ$, resulting in a total field of view of 0.53~sr.
    The system observes the soft X-ray band of 0.4--4~keV and localizes detected GRBs with arcminute accuracy.
    
    To cover the field of view of each EAGLE module and detect the focused X-rays, the focal-plane detector is required to have an imaging area of at least $55~\mathrm{mm} \times 55~\mathrm{mm}$. 
    Furthermore, high sensitivity is required to detect faint high-redshift GRBs.
    A time resolution better than 0.1~s is also required to capture the rapid temporal variability of GRBs.
    A pnCCD \cite{Lothar1987} developed by PNSensor GmbH is a promising candidate for the focal-plane detector that satisfies these requirements.
    
    However, the pnCCD outputs full-frame imaging data, and it is not feasible to downlink all raw image data to the ground due to telemetry limitations. 
    To reduce the data volume, X-ray event information must be extracted onboard the satellite. 
    In the space environment, charged-particles such as protons and electrons are incident on the detector, with particularly enhanced fluxes in high-latitude regions and in the South Atlantic Anomaly.
    Conventional X-ray event selection methods, such as the grade method \cite{Nakajima_2005}, may misidentify these charged particle events as X-ray events, leading to contamination of the data.
    Therefore, an onboard discrimination algorithm capable of distinguishing X-ray events from charged particle events and extracting only valid X-ray events is required.
    
    In this paper, we investigate and evaluate charged-particle rejection algorithm for implementation in the pnCCD electrical system of \textit{HiZ-GUNDAM}. 
    Section~2 describes the pnCCD required for EAGLE and our drive and readout system developed using a prototype small-format pnCCD. 
    Section~3 introduces the X-ray event extraction algorithms under consideration. 
    Section~4 presents experimental evaluations performed to verify their effectiveness. 
    Finally, Section~5 discusses the performance of the algorithms and their applicability to the \textit{HiZ-GUNDAM} mission.

\section{pnCCD and the readout system}
\label{sec2:pnCCD}
    The pnCCD developed by PNSensor GmbH has been employed in space missions such as \textit{eROSITA} \cite{Meidinger2010} and \textit{SVOM} \cite{Mercier2018}, demonstrating its flight heritage. 
    It provides a quantum efficiency exceeding 90\% in the soft X-ray band of 0.3--10~keV without optical blocking filters.
    In addition, the pnCCD is equipped with an analog readout ASIC called CAMEX \cite{CAMEX_Herrmann_2008}, which performs parallel readout over multiple channels. 
    This architecture enables high frame rates exceeding 10~fps, which are not achievable with conventional CCDs. 
    Therefore, the pnCCD is a strong candidate for the EAGLE focal-plane detector. 
    The requirements for the flight-model pnCCD are summarized in Table~\ref{tab:pnCCD_spec}.

\begin{figure}[htb]
        \centering
        \includegraphics[width=80mm]{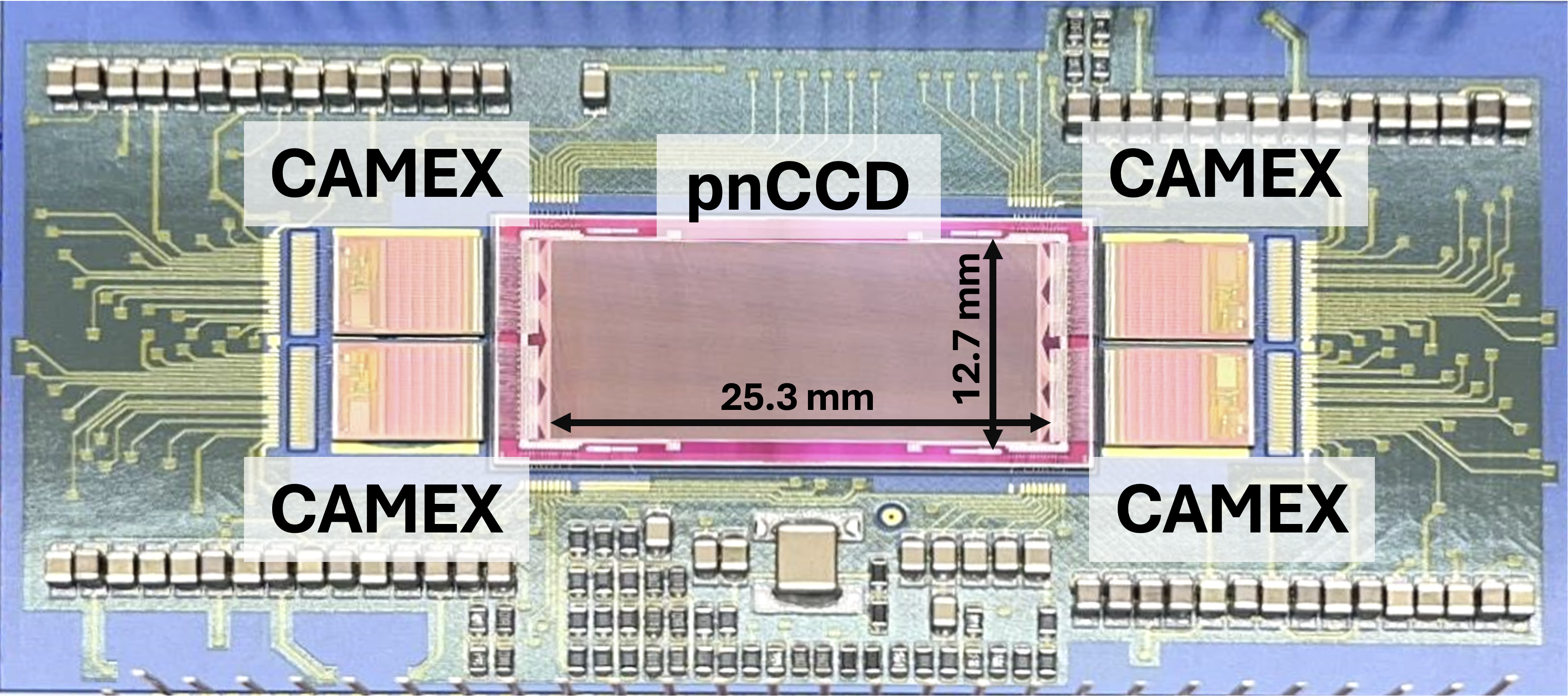}
        \caption{Photograph of the small pnCCD used in this study. The
        pnCCD sensor is mounted at the center of the chip, with two CAMEXs mounted at
        each end. }
        \label{fig:small_ccd}
    \end{figure}

    To verify the X-ray performance and evaluate onboard event-selection algorithms, we use a small breadboard model pnCCD (Figure~\ref{fig:small_ccd}). 
    The specifications of the small pnCCD are also listed in Table~\ref{tab:pnCCD_spec}. 
    The device is equipped with four CAMEX ASICs, each reading out one quarter of the imaging area.
    Each CAMEX has 132 parallel channels, enabling a maximum frame rate of 1300~fps.
    
    \begin{table}[htb]
        \centering
        \caption{Specifications of the pnCCD image sensor.}
        \begin{tabular}{lll}
            \toprule 
            Parameters  & Small pnCCD                                             & Flight-model pnCCD                \\
            \midrule 
            Sensitive Area   & $12.7 \times 25.3\ \rm{mm^2}$                         & $55 \times 55\ \rm{mm^2}$ \\
            Pixel size       & $132\ \rm{\upmu m}$                                   & $\sim 100\ \rm{\upmu m}$  \\
            Number of pixels & $96 \times 192$                                       & $512 \times 512$          \\
            Max frame rate   & $\sim$ 1300\ $\rm{fps}$                                       & $\sim$ 80\ $\rm{fps}$     \\
            Sensor type      & \multicolumn{2}{c}{Back-illuminated}                   \\
            Energy coverage  & \multicolumn{2}{c}{0.25 -- 15 keV}            \\
            Depletion layer  & \multicolumn{2}{c}{$450\ \rm{\upmu m}$ fully depleted} \\
            \bottomrule
        \end{tabular}
        \label{tab:pnCCD_spec}
    \end{table}
    
\begin{figure}[htb]
        \centering
        \includegraphics[width=80mm]{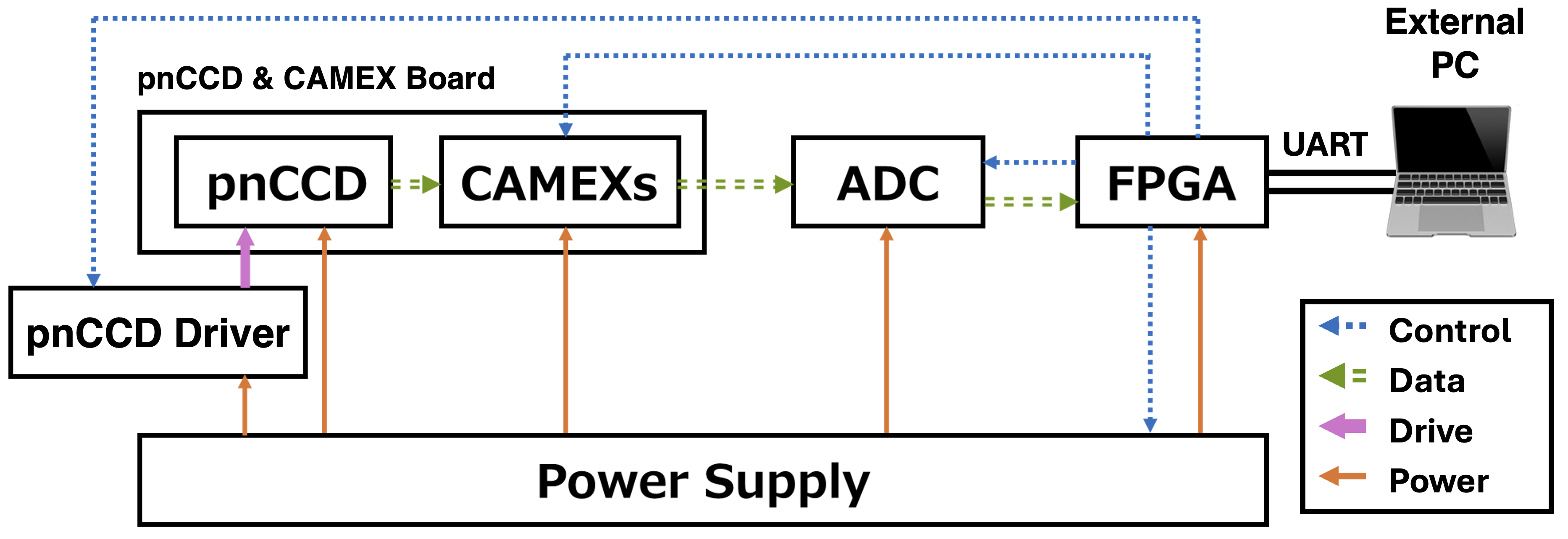}
        \caption{Block diagram of the developed pnCCD readout system.}
        \label{fig:pnCCD_diagram}
    \end{figure}

    For satellite implementation, we have developed a drive and readout system \cite{HCS2024, RK2024}.  
    The system consists of multiple electronic boards, each responsible for specific functions.
    Figure~\ref{fig:pnCCD_diagram} shows a block diagram of the developed readout electronics. 
    To achieve high-speed operation, we employ a field-programmable gate array (FPGA). 
    The FPGA generates the drive signals for the pnCCD and CAMEX. 
    The analog signals output from CAMEX are digitized and transferred back to the FPGA, where the data are temporarily stored before being transmitted to an external PC.
    
\section{X-ray event extraction and charged-particle event removal algorithms}
\label{sec3:Methods}
    The expected X-ray photon flux from high-redshift GRBs is relatively low. 
    Considering typical GRB brightness, the effective area of the Lobster Eye Optics, and the detection efficiency of the pnCCD, the expected detection count rate for high-redshift GRBs with EAGLE is approximately 0.4--2.0~counts~s$^{-1}$ per pnCCD.
    Therefore, the number of X-ray events recorded by the pnCCD per unit time is limited.
    Because of telemetry limitations, only event information rather than full-frame image data can be transmitted to the ground.
    Thus, X-ray event information, including position and energy, must be extracted onboard from the image data.
    
    In the space environment, charged particles such as protons and electrons interact with the detector and generate signals. 
    These charged-particle events constitute a background component in astronomical observations and can be misidentified as X-ray events. 
    Such misidentification can also lead to false alerts in GRB detection systems.
    To ensure efficient GRB detection and suppress background contamination, charged-particle events must be discriminated and removed before data transmission.

    For onboard event extraction, we consider both conventional grade methods used in previous X-ray astronomy missions and a newly developed CNN-based event selection method \cite{HCShen2023}.
    We evaluate and compare these methods to identify the most suitable approach for implementation in the pnCCD system under onboard resource constraints.
    In this section, we describe the event-classification methods used to distinguish X-ray events from charged-particle events.

\subsection{Grade Method}
\begin{figure}[htb]
        \centering
        \includegraphics[width=80mm]{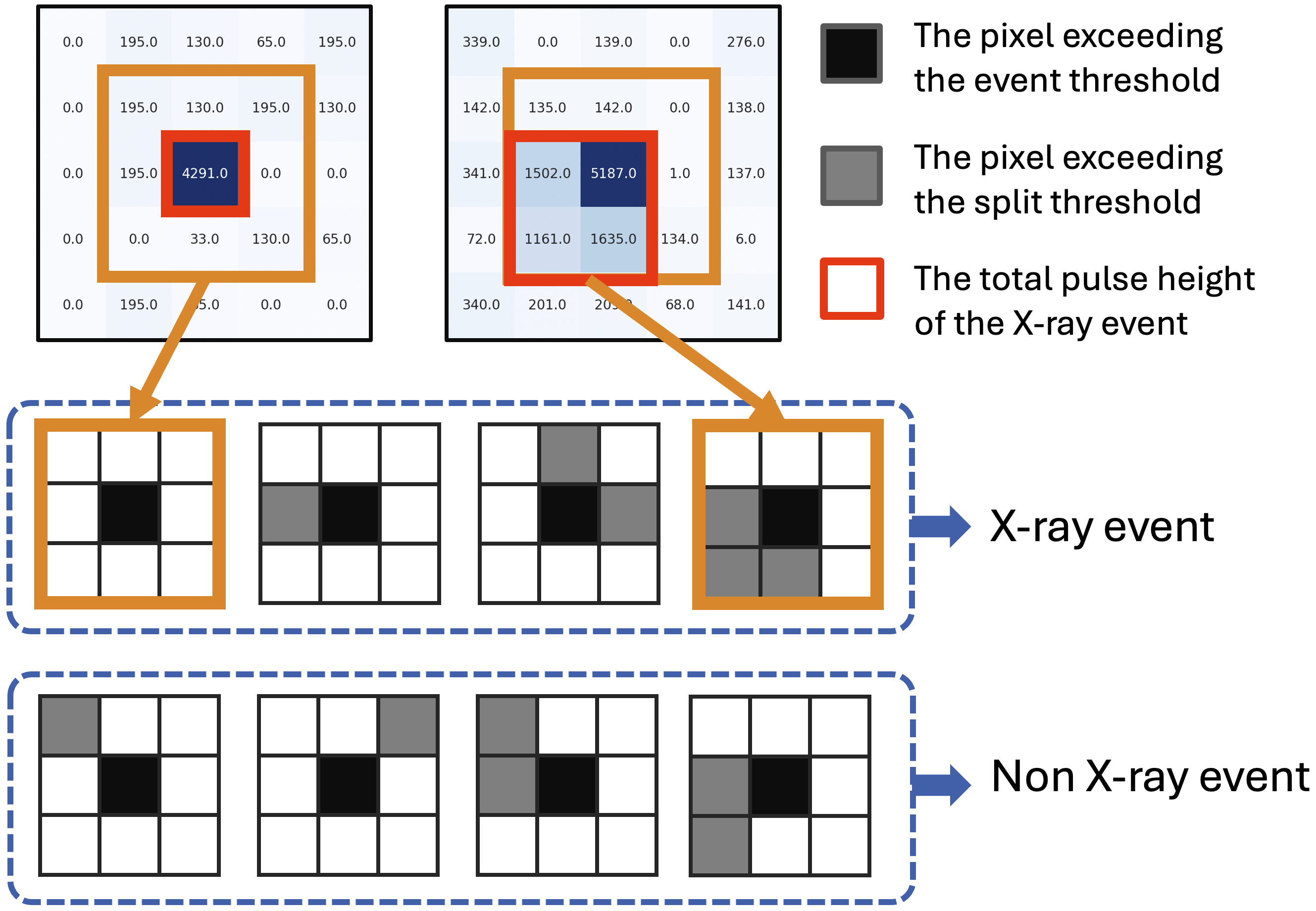}
        \caption{Grade method. Values in the pixels show the analog-to-digital units (ADU). The event pixel exceeds the event threshold, while surrounding pixels exceeding the split threshold are included for energy reconstruction. Representative pixel patterns used to classify X-ray and non–X-ray events are shown in the lower panels.}
        \label{fig:Grade}
    \end{figure}

    The grade method is widely used in X-ray astronomy to extract event information from CCD image data. 
    In this method, three thresholds are defined: an event threshold, a split threshold, and a particle threshold. 
    The event threshold identifies candidate event pixels corresponding to local charge maxima produced by incident radiation.
    The split threshold is applied to surrounding pixels (typically within a $3\times3$ or $5\times5$ region) to detect multi-pixel charge-sharing events. 
    Based on the pattern of pixels exceeding the event and split thresholds, events are classified as X-ray or charged-particle events.
    Examples of representative grade patterns used for X-ray and charged-particle classification are shown in Fig.~\ref{fig:Grade}.
    The pixel values exceeding the event and split thresholds are summed to obtain the total deposited energy. 
    Events whose total signal exceeds the particle threshold are classified as charged-particle events and rejected.
    
    In the ASCA mission \cite{YTanaka1994}, the event recognition region was limited to $3\times3$ pixels due to onboard computational constraints. 
    From the Suzaku mission \cite{KMitsuda2007} onward, the recognition region was expanded to $5\times5$ pixels, allowing events extending beyond the $3\times3$ region to be identified and rejected as charged-particle events.

\subsection{CNN Method} 
    Recent advances in machine learning have significantly improved image recognition techniques. 
    Since charged-particle events typically produce more spatially extended signal patterns than X-ray events, we investigated a machine-learning-based X-ray event classification method~\cite{HCShen2023}.
    
    A convolutional neural network (CNN) was employed to discriminate between X-ray and charged-particle events. 
    For training, clipped images of $5\times5$ pixels centered on candidate events were prepared, including both X-ray and charged-particle samples. 
    The adopted CNN architecture employed in this study consists of two convolutional layers with $3\times3$ kernels, followed by dropout regularization and two fully connected layers for classification.
    The CNN model was trained using a sparse categorical cross-entropy loss function and the Adam optimizer, with a batch size of 32 for 250 epochs.

    To improve robustness against variations in track orientation, data augmentation was applied to the training dataset using rotations by 90°, 180°, and 270°, as well as horizontal and vertical reflections.
    Because charged-particle tracks have no preferred orientation, these transformations preserve the physical characteristics of the events while increasing the diversity of the training samples.

    We evaluated several CNN configurations, including a two-layer CNN with a $5\times5$ input window, a three-layer CNN with a $5\times5$ input window, and a two-layer CNN with a $9\times9$ input window. 
    The two-layer CNN with a $5\times5$ input window provided the best balance between classification accuracy and model complexity. 
    Considering future onboard implementation constraints, this model was adopted for the present study.
    The performance comparison of these configurations is presented in Section~\ref{sec4_1:Experiments} .
    
\begin{figure}[htb]
        \centering
        \includegraphics[width=65mm]{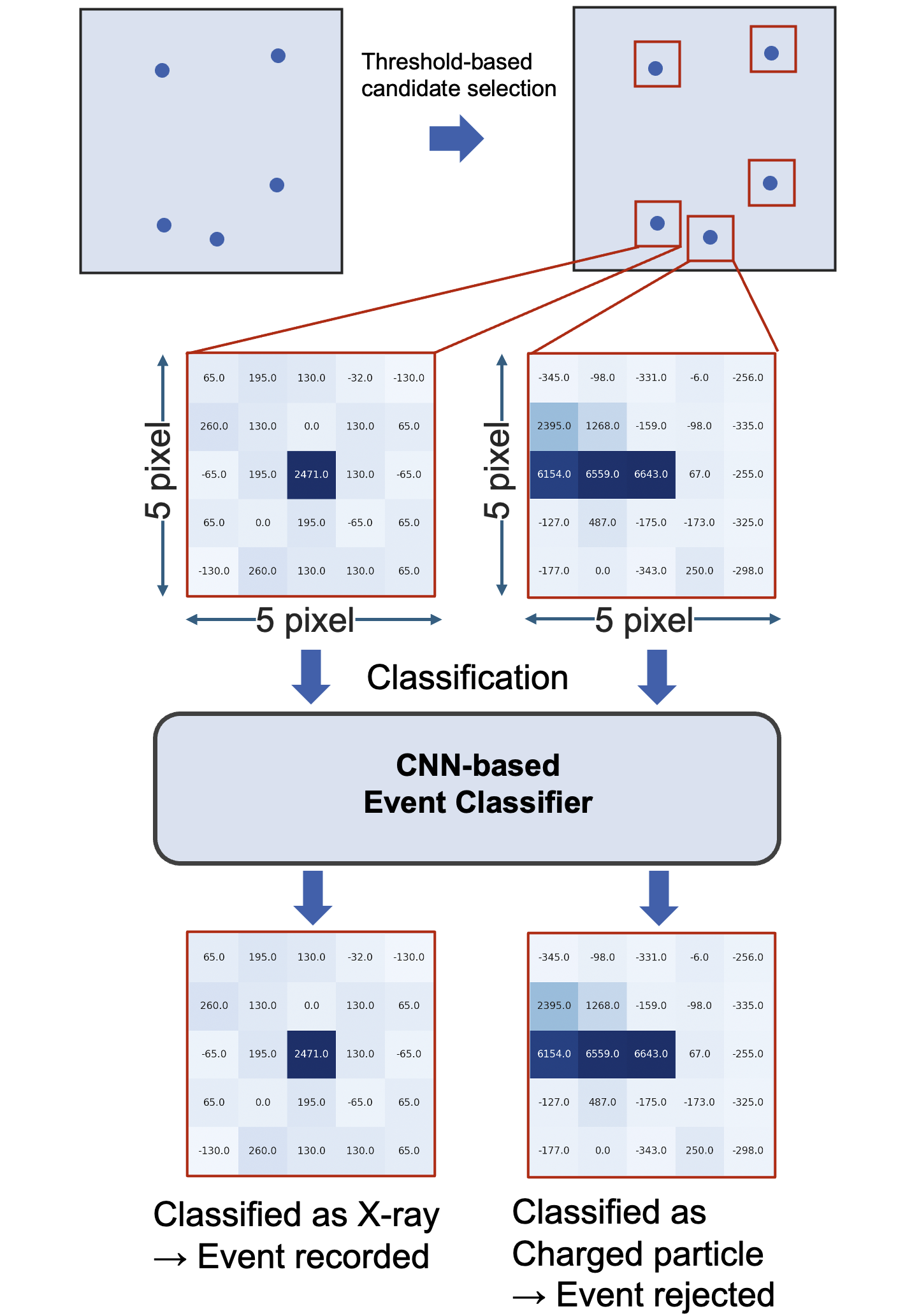}
        \caption{Workflow of the CNN-based X-ray event selection method. 
        Candidate center pixels are first identified using a threshold. 
        A $5\times5$ pixel patch around each candidate is extracted and classified by a CNN event classifier. 
        Events identified as X-rays are recorded, while charged-particle events are rejected. Values in the pixels show the analog-to-digital units (ADU).}
        \label{fig:MLM}
    \end{figure}

    The event selection procedure is shown in Fig.~\ref{fig:MLM}. 
    First, candidate center pixels are identified using a predefined threshold. 
    A $5\times5$ pixel region surrounding each candidate is then extracted and fed to the trained CNN model. 
    If the event is classified as an X-ray, its position and energy information are recorded; otherwise, the event is rejected.
    
    In the following section, the performance of the grade and CNN methods is evaluated experimentally.

\section{Experimental Verification}
\label{sec4:Experiments}

    To evaluate the practical applicability of the grade method and the CNN method, we performed verification experiments using our developed pnCCD drive and readout system.
    The pnCCD was irradiated with X-rays and charged particles, and the acquired image data were processed using both the grade method and the CNN method. 
    Based on the extracted events, the charged-particle rejection performance and X-ray acceptance rate were investigated.
    In this section, we describe the experimental setup, data acquisition conditions, and the evaluation results.
    
\subsection{Experimental Setup} 
\label{sec4_1:Experiments}
\begin{figure}[htbp]
        \centering
        \includegraphics[width=65mm]{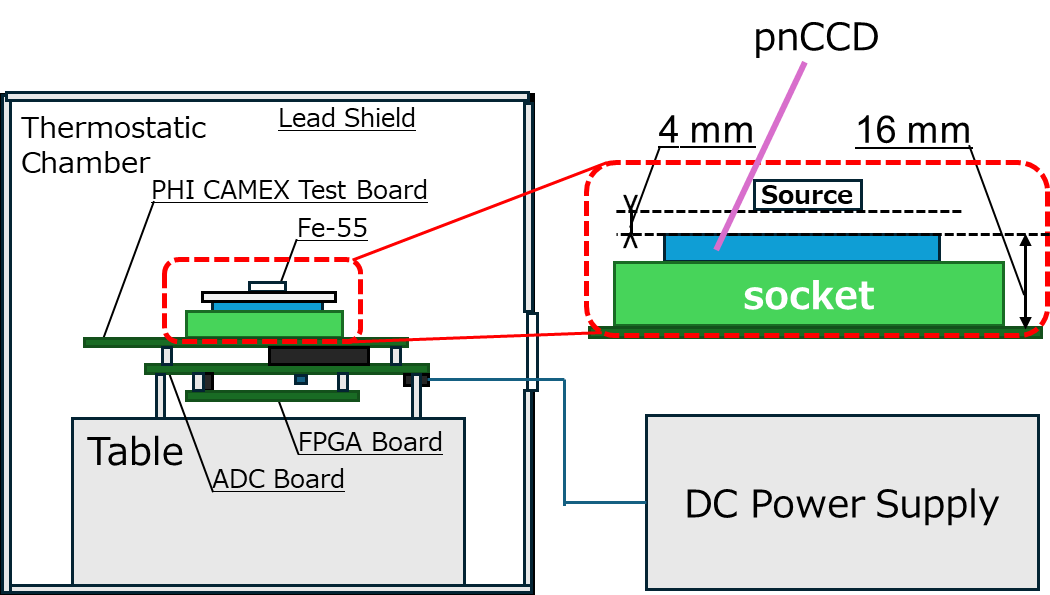}
        \caption{The experimental setup. The pnCCD is mounted on a socket and connected to FPGA and ADC boards, with power supplies. The system was placed in a thermostatic chamber. The radioactive sources were positioned at a distance of 4~mm from the pnCCD.}
        \label{fig:setup}
    \end{figure}

    Figure~\ref{fig:setup} shows the experimental setup used in this study. 
    The pnCCD was placed in a thermostatic chamber and its temperature was stabilized at 0~$^\circ$C during the measurements.
    
    An $^{55}$Fe radioactive source, emitting Mn K$\alpha$ (5.9~keV) and K$\beta$ (6.5~keV) X-rays, was used as the X-ray source. 
    Since electrons are expected to be a dominant background source in low-Earth orbit environments relevant to the mission, a $^{90}$Sr radioactive source emitting $\beta$-rays with a maximum energy of 0.546~MeV was employed as the charged-particle source.  
    The sources were positioned at a distance of 4~mm from the pnCCD. 
    The X-ray and $\beta$-ray measurements were performed separately.
    
    Table~\ref{tab:data_conditions} summarizes the experimental conditions.
    To construct the CNN model, image data were acquired with different exposure times: 20~ms per frame for X-ray irradiation and 25~ms per frame for $\beta$-ray irradiation.
    A total of 20 frames were recorded for X-ray irradiation and 100 frames for $\beta$-ray irradiation.
    These acquisition conditions were chosen to obtain approximately balanced numbers of X-ray and $\beta$-ray events for training, resulting in about 800 events in each class.
    
    The event dataset was first divided into training (80\%) and test (20\%) subsets using a stratified random split to preserve the class balance.
    Data augmentation was applied only to the training subset.
    To assess the robustness and generalization performance of the CNN model, five-fold stratified cross-validation was performed on the training dataset.
    The held-out test subset was used to evaluate the classification performance after training.

    To determine an appropriate CNN architecture, several model configurations were evaluated using the same cross-validation procedure. 
    Table~\ref{tab:cnn_comparison} summarizes the cross-validation and test performance of the investigated CNN configurations.
    The two-layer CNN with a $5\times5$ input window achieved the highest mean cross-validation accuracy and the smallest model complexity among the evaluated configurations.
    Increasing the network depth to three convolutional layers or expanding the input window to $9\times9$ pixels did not improve the classification performance. 
    Therefore, the two-layer CNN with a $5\times5$ input window was adopted for the subsequent analysis.
    The small variation observed across the five cross-validation folds (standard deviation of 0.23\% for the selected model) suggests that the CNN performance is robust against variations in the training dataset and is not strongly dependent on a particular train/validation split.
    
    For performance evaluation of the grade method and the CNN method, an independent dataset was acquired separately from the training data. 
    For this validation dataset, 100 frames were recorded for both X-ray and $\beta$-ray measurements under the same exposure condition. 
    The training and validation datasets were kept separate to avoid overfitting.
    
    \begin{table}[htbp]
        \centering
        \caption{Experimental conditions for data acquisition using X-ray and
        $\beta$-ray sources.}
        \label{tab:data_conditions}
        \begin{tabular}{lcc}
            \toprule Parameter          & X-ray source                & $\beta$-ray source    \\
            \midrule Source             & $^{55}\text{Fe}$            & $^{90}\text{Sr}$      \\
            Temperature                 & 0 \si{\celsius}             & 0 \si{\celsius}       \\
            Integration time per frame  & 20 \si{ms}                  & 25 \si{ms}            \\
            Data sets for training      & 20 frames                   & 100 frames            \\
            Data sets for verification  & 100 frames                  & 100 frames            \\
            
            \bottomrule
        \end{tabular}
    \end{table}

    \begin{table}[htbp]
        \centering
        \caption{Performance comparison of different CNN configurations.}
        \label{tab:cnn_comparison}
        \begin{tabular}{llll}
            \toprule Model                   & Mean CV    & Std. [\%]    & Test  \\
                                             & Acc. [\%]  &              & Acc. [\%] \\
            \midrule $5\times5$ 2-layer CNN  & 99.43      & 0.23          & 99.04     \\ 
            $5\times5$ 3-layer CNN           & 99.37      & 0.36          & 99.04     \\
            $9\times9$ 2-layer CNN           & 99.01      & 0.15          & 98.33     \\
            
            \bottomrule
        \end{tabular}

        \begin{tablenotes}
        \footnotesize
        \item Mean CV Acc.: Mean classification accuracy obtained from five-fold cross-validation.
        \item Std.: Standard deviation of the cross-validation accuracy.
        \item Test Acc.: Classification accuracy evaluated using the independent test dataset.
        \end{tablenotes}
        
    \end{table}
    
    Prior to event discrimination, dark-frame subtraction was performed to remove pixel offset and dark current contributions.
    The root mean square (RMS) noise level, defined as the standard deviation of pixel output values measured under dark conditions, was evaluated from the dark-subtracted frames.
    The event threshold was set to 5$\sigma$ of the RMS noise level to suppress spurious noise triggers, while the split threshold was set to 3$\sigma$ to identify charge-sharing signals in adjacent pixels while minimizing noise contamination.
    These values were chosen to balance X-ray detection efficiency and noise rejection performance.
    
    In this analysis, the event threshold and split threshold were set to 1000 and 600 ADU, corresponding to $\sim$1.4~keV and $\sim$0.8~keV.
    Although the target observation band of EAGLE is 0.4 -- 4~keV, the effective lower energy bound in this experiment is limited to $\sim$1.4~keV due to the relatively high readout noise of the current system. 
    This limitation arises from the experimental setup and does not reflect the intrinsic performance of the detector, which is expected to reach the mission requirement with an optimized readout system.
    The particle threshold was set to the maximum ADU value (16383), corresponding to the 14-bit dynamic range of the readout system.
    Events associated with penetrating muons are expected to be effectively rejected by this threshold, since such particles typically deposit on the order of 200~keV while traversing the 450~$\mu$m depletion layer, producing signals that exceed the dynamic range of the detector.

\subsection{Results} 
\begin{figure}[htbp]
        \centering
        \includegraphics[width=80mm]{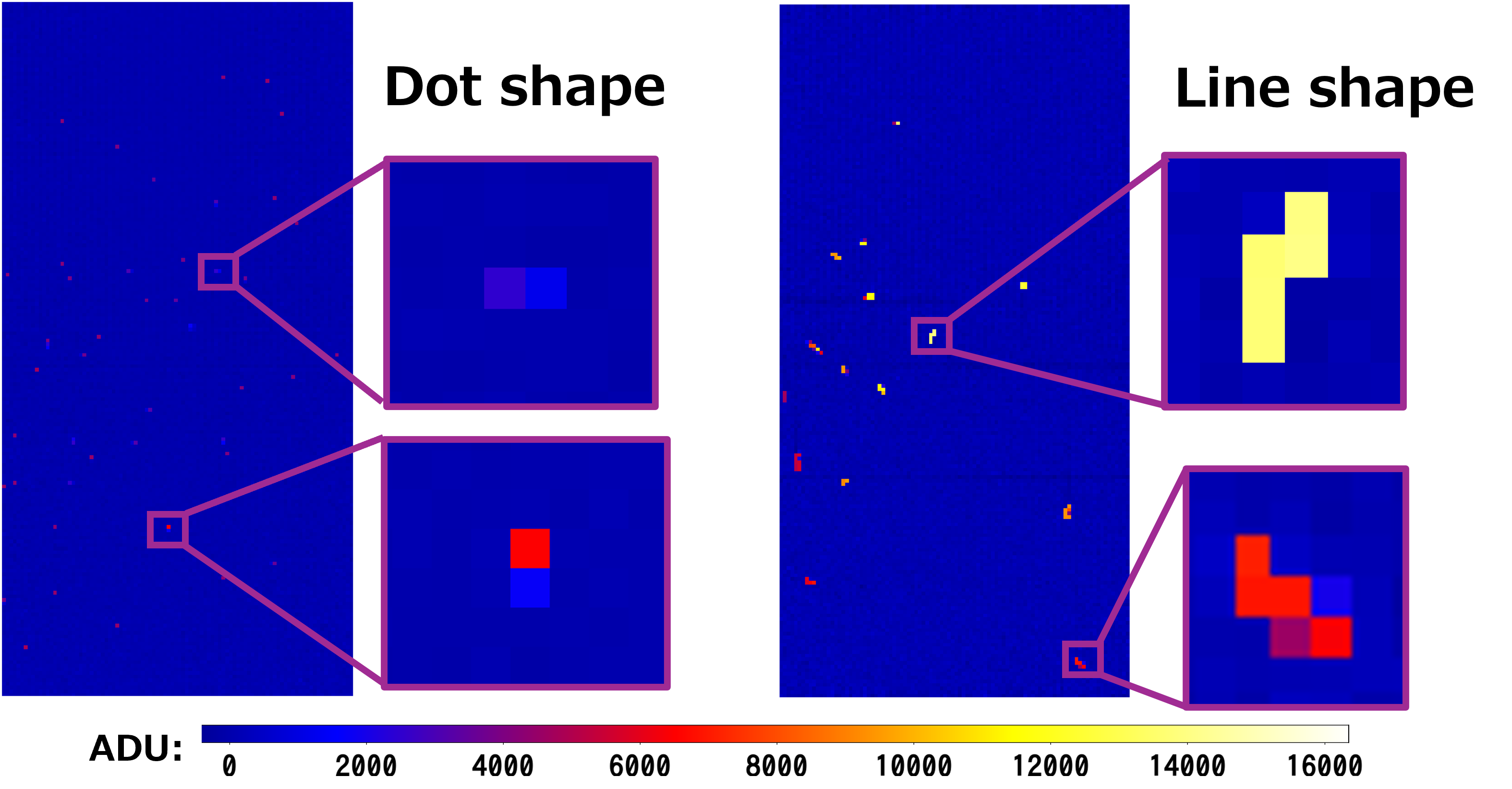}
        \caption{Representative X-ray (left) and $\beta$-ray (right) pnCCD images. X-ray events appear as compact, dot-like patterns, whereas $\beta$-ray events form extended, track-like structures. Insets show magnified examples. Pixel values are expressed in analog-to-digital units (ADU).}
        \label{fig:4_2_x_beta_frame}
    \end{figure}

    Figure~\ref{fig:4_2_x_beta_frame} shows representative images obtained under X-ray and $\beta$-ray irradiation.
    X-ray events exhibit spatially compact charge distributions, typically confined to a few pixels. 
    In contrast, $\beta$-ray events produce extended track-like patterns due to multiple scattering and continuous energy deposition within the silicon layer of the pnCCD.
    
\begin{figure}[htbp]
        \centering
        \includegraphics[width=82mm]{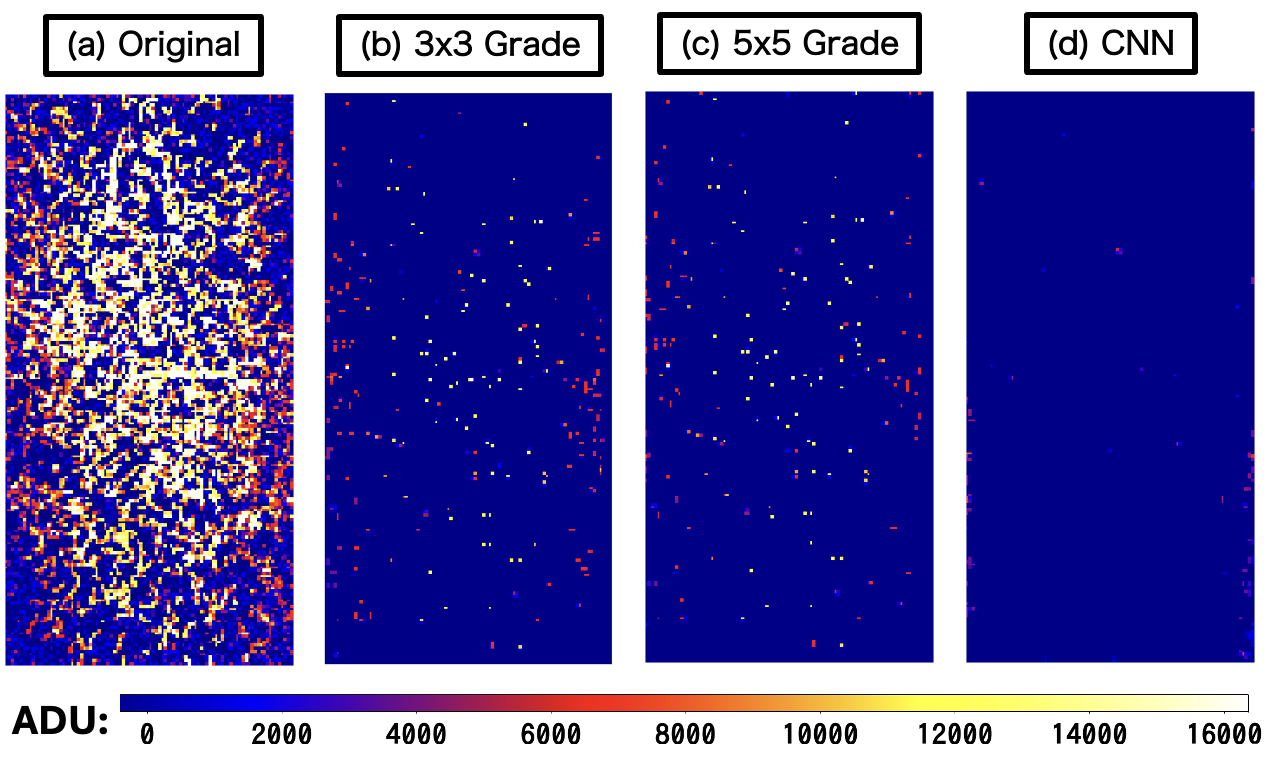}
        \caption{Comparison of $\beta$-ray event maps from pnCCD data acquired with a $^{90}\text{Sr}$ source: (a) Original image; (b) Events misclassified with the 3$\times$3 grade method; (c) Events misclassified with the 5$\times$5 grade method; (d) Events misclassified with the CNN method.}
        \label{fig:beta_countmaps}
    \end{figure}

    To evaluate the charged-particle rejection performance, the 100-frame $\beta$-ray validation dataset was processed using the $3\times3$ grade method, the $5\times5$ grade method, and the CNN method. 
    Event maps were generated based on the extracted event positions, as shown in Figure~\ref{fig:beta_countmaps}. 
    Since charged-particle events are background signals, events appearing in the event maps correspond to $\beta$-ray events misidentified as X-ray events. 
    As shown in Figure~\ref{fig:beta_countmaps}, misclassified events remain when using the $3\times3$ and $5\times5$ grade methods, whereas the CNN method suppresses most of the $\beta$-ray events.

    The misclassification rate was calculated as the fraction of $\beta$-ray events that were classified as X-ray events.
    The misclassification rates are 11.9\% for the $3\times3$ grade method and 10.2\% for the $5\times5$ grade method, while the CNN method achieves a misclassification rate of 3.1\%. 
    This corresponds to an approximately four-fold reduction in the misclassification rate compared with the conventional grade methods.
    
\begin{figure}[htbp]
        \centering
        \includegraphics[width=82mm]{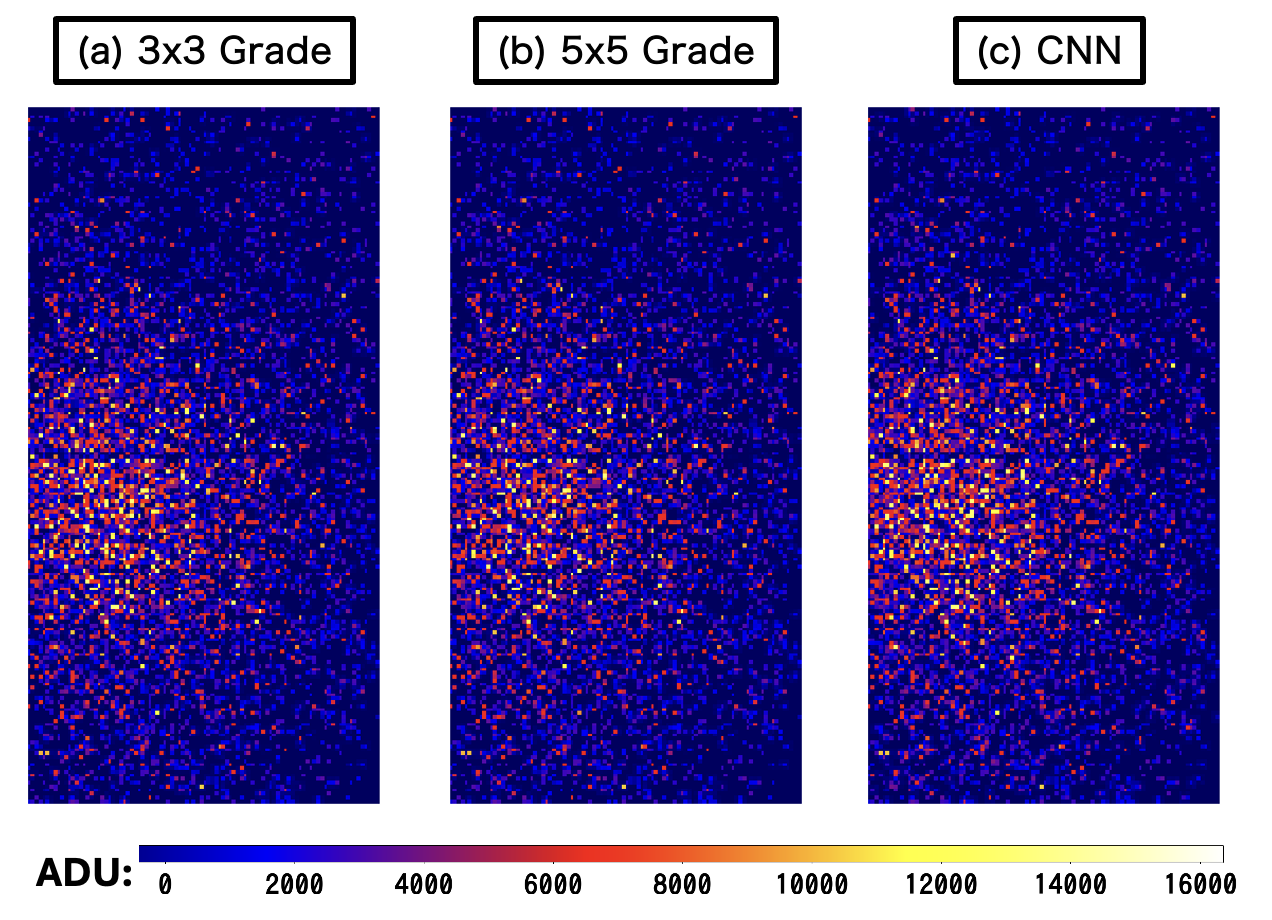}
        \caption{Comparison of X-ray count maps from pnCCD data acquired with a $^{55}\text{Fe}$ source: (a) Events
        extracted by the 3$\times$3 grade method; (b) Events extracted by the 5$\times$5
        grade method; (c) Events extracted by the CNN method.}
        \label{fig:xray_countmaps}
    \end{figure}

    In addition to charged-particle rejection, accurate X-ray event extraction is equally important. 
    To evaluate the X-ray acceptance performance, the 100-frame X-ray validation dataset obtained under $^{55}$Fe irradiation was processed using the same three methods ($3\times3$ grade method, $5\times5$ grade method and CNN method). 
    The extracted X-ray event maps are shown in Figure~\ref{fig:xray_countmaps}. 
    
    The X-ray acceptance rate is defined as the ratio of the number of events classified as X-ray events to the total number of detected events, where the total includes both events classified as X-rays and those rejected as charged-particle events.
    This metric represents the fraction of events in the $^{55}$Fe validation dataset that are accepted as X-ray events after classification and does not represent the absolute detection efficiency of incident X-rays.
    To suppress contamination from cosmic-ray muons and other high-energy charged-particle events, events with pulse heights exceeding the maximum ADU value (16383) were excluded from the analysis.
    In addition, the event threshold was set to 5$\sigma$ above the readout-noise level. 
    Inspection of the resulting $^{55}$Fe spectra confirmed that no significant residual noise component remained in the analyzed dataset. 
    All events in the $^{55}$Fe irradiation dataset that passed these selection criteria were therefore used as the reference dataset for evaluating the X-ray acceptance rate. 
    Consequently, contamination from noise and high-energy charged-particle events is expected to be minimal.
    
    The X-ray acceptance rates are 97.8\% for the $3\times3$ grade method, 91.9\% for the $5\times5$ grade method, and 98.1\% for the CNN method. 
    The lower X-ray acceptance rate of the $5\times5$ grade method is attributable to pile-up events present in the $^{55}$Fe validation dataset. 
    In such events, multiple X-ray events are recorded within a $5\times5$ pixel event window, resulting in charge distributions that do not satisfy the $5\times5$ grade selection criteria and are therefore rejected.
    These results demonstrate that the CNN method maintains a high X-ray acceptance rate while substantially improving charged-particle rejection performance.

    \begin{table*}[htbp]
        \centering
        \begin{threeparttable}
        \caption{Quantitative comparison of the charged-particle misclassification rate and X-ray acceptance rate for \\ different discrimination methods.}
        \begin{tabular}{lrrrrr}
            \toprule
             \multirow{2}{*}{\shortstack[l]{Sensor}} & \multirow{2}{*}{\shortstack[l]{Depletion layer}} & \multirow{2}{*}{\shortstack[l]{Method}} & \multicolumn{2}{c}{Misclassification rate}& \multirow{2}{*}{\shortstack[l]{X-ray acceptance rate}} \\
             &   &  & $1.4-23~\rm{keV}$ & $1.4-4.0~\rm{keV}$ & \\
            \hline
            \multirow{3}{*}{\shortstack[l]{pnCCD (this work)}}
             & \multirow{3}{*}{450~$\rm{\upmu m}$}      & 3$\times$3 grade & $11.9 \pm 0.7$\% & $0.4 \pm 0.1$\% & $97.8 \pm 0.2$\% \\
             &                                          & 5$\times$5 grade & $10.2 \pm 0.7$\% & $0.5 \pm 0.2$\% & $91.9 \pm 0.3$\% \\
             &                                          & CNN              & $3.1 \pm 0.4$\% & $0.7 \pm 0.2$\% & $98.1 \pm 0.2$\% \\
            \hline
            \multirow{3}{*}{CMOS \cite{HCShen2023}}
             & \multirow{2}{*}{$\sim10$~$\rm{\upmu m}$} & 3$\times$3 grade & \multicolumn{2}{c}{$89.3 \pm 1.0$\% ($0.4-7.5~\rm{keV}$)} &  $82.1 \pm 0.1$\%  \\
             &                                          & CNN              & \multicolumn{2}{c}{$18.2 \pm 1.2$\% ($0.4-7.5~\rm{keV}$)} &  $80.4 \pm 0.1$\%  \\
            \hline
        \end{tabular}
        \label{tab:performance_comparison}
        \begin{tablenotes}
            \footnotesize
            \item{*}The uncertainties of the misclassification rate and X-ray acceptance rate were estimated assuming Poisson statistics.
        \end{tablenotes}
        \end{threeparttable}
    \end{table*}

    Table~\ref{tab:performance_comparison} summarizes the misclassification rate and the X-ray acceptance rate obtained in this work for each method.
    The misclassification rate is evaluated both over the full energy range (up to the ADC saturation level of 16383 ADU, corresponding to $\sim$23~keV) and within the 1.4 -- 4~keV band.
    Although the primary observation band of \textit{HiZ-GUNDAM} is 0.4 -- 4~keV, the lower bound of the evaluated energy range is limited to $\sim$1.4~keV in this experiment due to the readout noise level, as described above.

\begin{figure}[htbp]
        \centering
        \includegraphics[width=80mm]{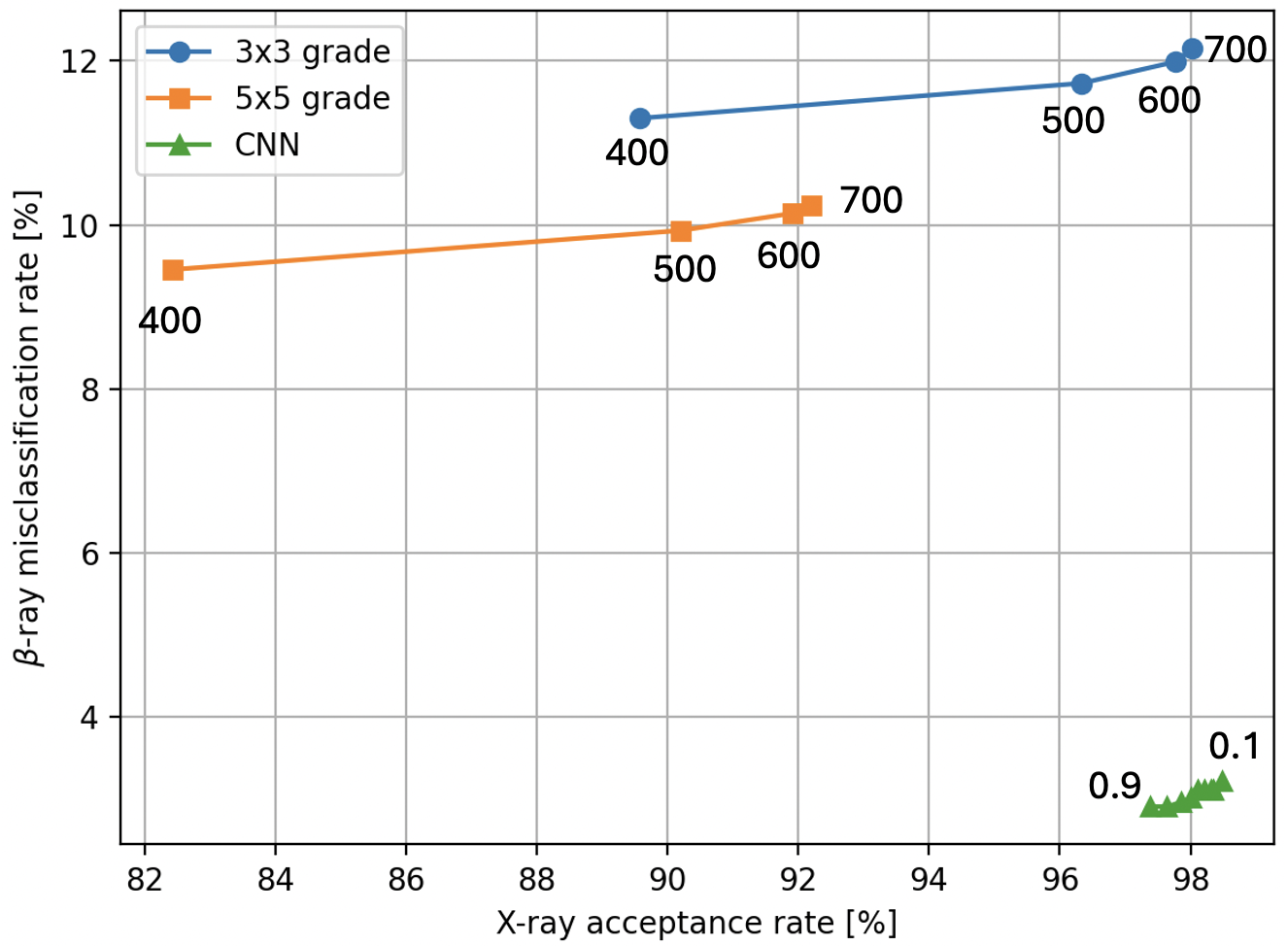}
        \caption{Trade-off between the X-ray acceptance rate and the $\beta$-ray misclassification rate for the $3\times3$ grade method, the $5\times5$ grade method, and the CNN method.
        For the grade methods, the split threshold was varied from 400 to 800 ADU.
        For the CNN method, the decision threshold was varied using the softmax probability of the X-ray class from 0.1 to 0.9.}
        \label{fig:roc_curve}
    \end{figure}

    To evaluate the dependence of the classification performance on the selected operating point, the decision thresholds of both the grade and CNN methods were varied.
    Figure~\ref{fig:roc_curve} shows the resulting trade-off between the X-ray acceptance rate and the $\beta$-ray misclassification rate.
    For the grade methods, the split threshold was varied from 400 to 800 ADU, while for the CNN method the decision threshold was varied using the softmax probability of the X-ray class.
    As shown in Figure~\ref{fig:roc_curve}, the CNN method maintains a substantially lower $\beta$-ray misclassification rate than the grade methods while preserving a high X-ray acceptance rate over a wide range of operating points.
    The classification performance of the CNN method shows only a weak dependence on the selected decision threshold within the investigated range.

\begin{figure}[htbp]
        \centering
        \includegraphics[width=80mm]{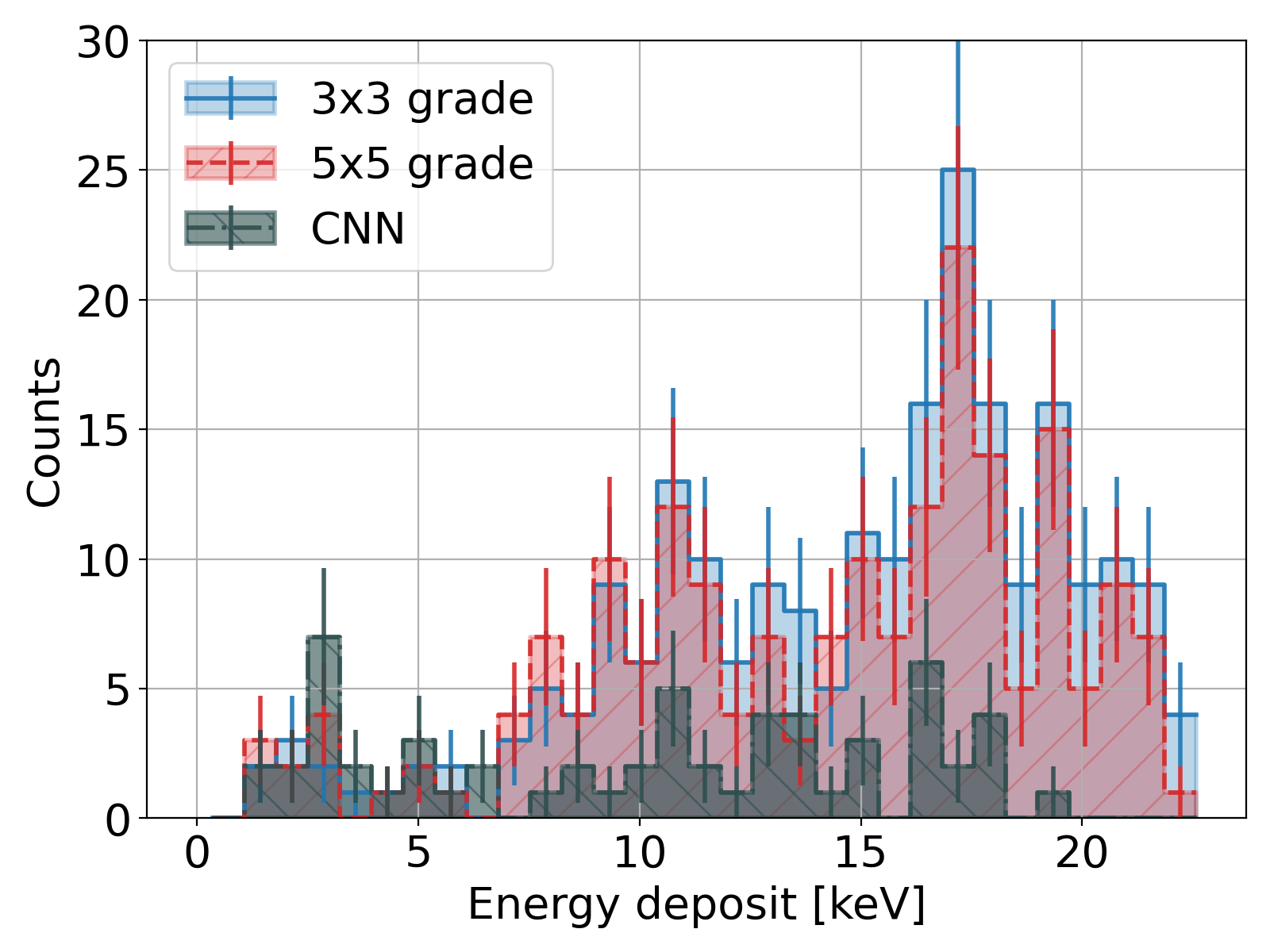}
        \caption{Energy spectra of $\beta$-ray events misclassified as X-ray events (false X-ray events) obtained with the $3\times3$ grade method, the $5\times5$ grade method, and the CNN method on pnCCD image data irradiated with $\beta$-rays. Error bars represent statistical uncertainties assuming Poisson statistics, given by the square root of the counts in each bin.}
        \label{fig:MCEvent_spectrum}
    \end{figure}

    To further investigate the characteristics of the remaining misclassified events, the energy spectra of false X-ray events were examined.
    Figure~\ref{fig:MCEvent_spectrum} shows the energy spectra of $\beta$-ray events misclassified as X-ray events (false X-ray events) for the $3\times3$ grade method, the $5\times5$ grade method, and the CNN method. 
    The grade methods yield a larger number of false events, and the resulting false event spectrum is biased toward higher deposited energies. 
    In contrast, the CNN method suppresses false events over the entire range and particularly reduces those at higher deposited energies.
    
\section{Discussion}
\label{sec:Discussion}
    The experimental results demonstrate that the CNN method significantly improves charged-particle rejection performance while maintaining a high X-ray acceptance rate. 
    In this section, we discuss the reasons for this performance difference.

\subsection{Interpretation of Discrimination Performance} 
\begin{figure}[htbp]
        \centering
        \includegraphics[width=80mm]{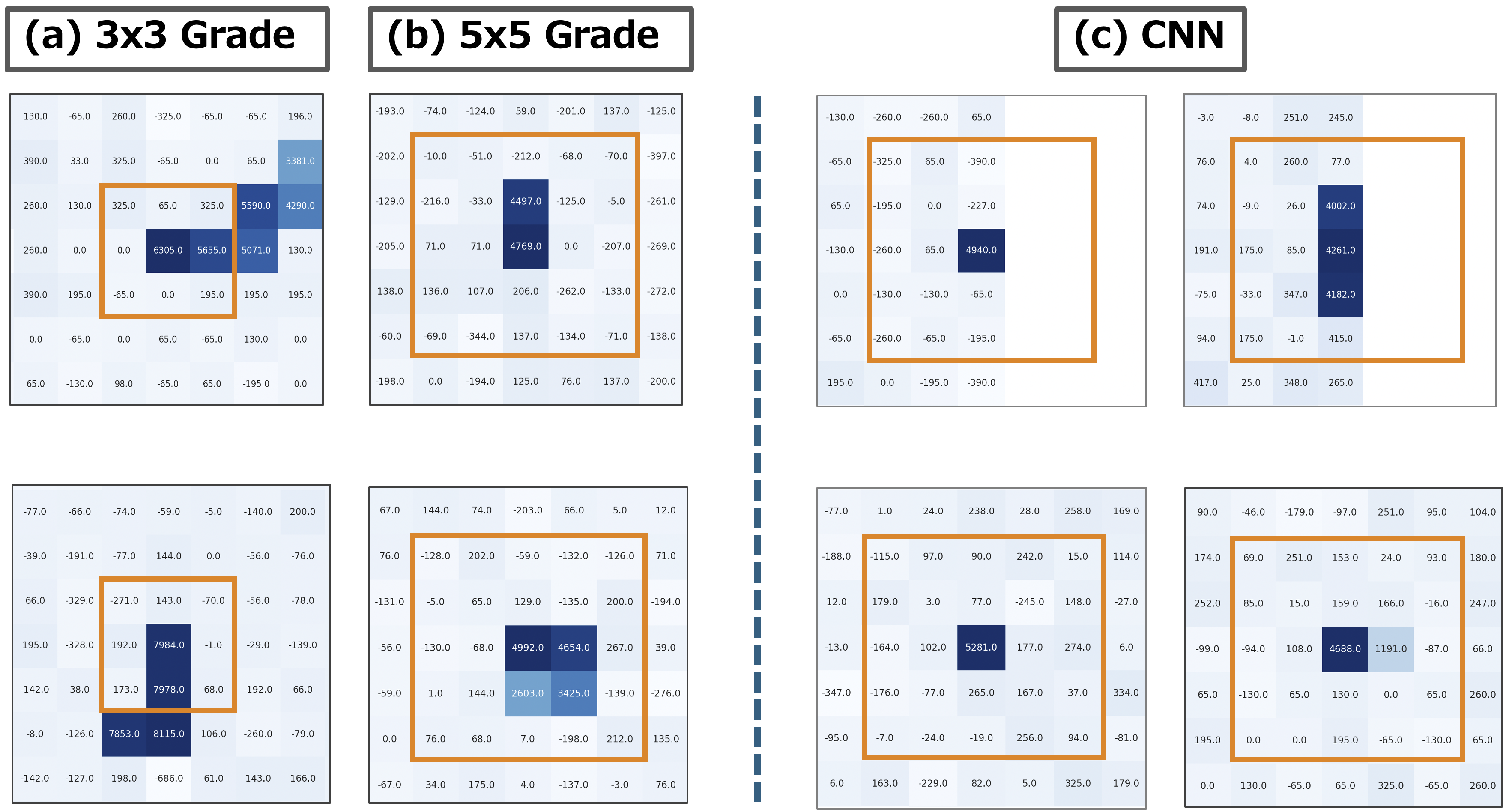}
        \caption{Representative $\beta$-ray events misclassified as X-ray events by (a) the $3\times3$ grade method, (b) the $5\times5$ grade method, and (c) the CNN method. The orange boxes indicate the pixel regions used for discrimination. The grade methods misidentify extended or track-like charge distributions, whereas the CNN method suppresses most such events, with only a few compact or edge events remaining misclassified.}
        \label{fig:MC_events}
    \end{figure}

    Figure~\ref{fig:MC_events} shows representative examples of $\beta$-ray events misclassified as X-ray events by the $3\times3$ grade method, the $5\times5$ grade method, and the CNN method.
    
    The $3\times3$ grade method performs event discrimination within a limited pixel region. 
    As a result, extended charge distributions that exceed the $3\times3$ window are frequently misclassified as X-ray events. 
    The $5\times5$ grade method expands the recognition region and successfully removes some of these extended events. 
    However, events confined within the $3\times3$ region but exhibiting similarly high signals in adjacent pixels may still be misclassified.
    Such patterns arise when energetic electrons traverse the silicon layer and deposit comparable amounts of energy across neighboring pixels. 
    These charge distributions are characteristic of charged-particle tracks.
    In contrast, X-ray photons generate charge clouds that spread radially from the interaction point. 
    For multi-pixel X-ray events, the signal in surrounding pixels is significantly lower than that of the central pixel.
    As a result, discrimination based on fixed pixel windows becomes challenging for such events.
    The CNN method evaluates the full $5\times5$ pixel pattern and captures subtle spatial correlations among pixel signals.
    This enables improved discrimination of track-like patterns that cannot be fully captured by fixed-window, rule-based grade methods.

    The energy distribution of misclassified events provides further insight into the performance difference between the methods. 
    As shown in Figure~\ref{fig:MCEvent_spectrum}, false X-ray events obtained with the grade methods are not uniformly distributed in energy but are biased toward higher deposited energies. 
    This suggests that the misclassified events are not extended track-like events, but rather compact charge deposits with relatively large signal amplitudes, which can arise from localized energy deposition by charged particles in the silicon layer.
    Since grade methods rely on simple thresholding and limited spatial patterns, these compact, high-amplitude events can be misidentified as X-ray events. 
    In contrast, the CNN method effectively suppresses such false events at higher deposited energies by utilizing both the spatial structure of charge distributions and the relative signal differences among pixels, enabling discrimination between compact X-ray events and charged-particle tracks.

    To evaluate the impact of the fixed $5\times5$ input window, the spatial extent of $\beta$-ray events was examined using $9\times9$ pixel images. 
    An event was classified as extending beyond the $5\times5$ window when at least one pixel outside the central $5\times5$ region exceeded the split threshold of 600 ADU (corresponding to 3$\sigma$ of the RMS noise level). 
    Using this criterion, 47.9\% of the $\beta$-ray events in the dataset were found to extend beyond the boundaries of the $5\times5$ window.
    Despite this, expanding the input size from $5\times5$ to $9\times9$ pixels resulted in a slight reduction in classification accuracy.
    This suggests that the charge-distribution features required for event discrimination are largely contained within the central $5\times5$ region, while the additional peripheral information contributes little to the classification performance.

\subsection{Influence of Detector Structure} 
    
    The detector structure also affects the spatial distribution and extent of charge signals over multiple pixels.
    The pnCCD used in this study has a thick depletion layer of 450~$\rm{\upmu m}$, allowing charged particles to traverse a substantial sensitive volume and deposit energy continuously along their trajectories. 
    This generally produces extended charge distributions and relatively large total signals, which enhances the effectiveness of spatial-extent criteria and particle-threshold-based rejection.
    
    However, the relatively large pixel size (132~$\rm{\upmu m}$) limits spatial sampling of complex charge tracks, sometimes compressing diverse morphologies into a small number of pixels. 
    Under such conditions, fixed-window grade methods may not fully capture the variability of charged-particle patterns, whereas the CNN method can learn more flexible spatial features.
    
    For comparison, detectors with much thinner depletion layers, such as CMOS sensors with thicknesses of order 10~$\rm{\upmu m}$, provide a significantly smaller sensitive volume. 
    Charged particles traversing such thin layers deposit energy over shorter path lengths, and the total deposited charge may approach that of soft X-ray events. 
    In such cases, discrimination based solely on signal may become less effective.
    Therefore, the effectiveness of different discrimination features, such as spatial extent and signal distribution, depends strongly on the detector structure.

    In this context, a previous study using a thin CMOS sensor (Gpixel GSENSE400BSI \cite{NO2021}, depletion thickness of approximately 10~$\rm{\upmu m}$) reported a misclassification rate of 18.2\% under $\beta$-ray irradiation ($^{210}$Bi, maximum energy 1.16~MeV) when applying a CNN method \cite{HCShen2023} (Table~\ref{tab:performance_comparison}, lower rows).
    This value is higher than that obtained with the pnCCD (3.1\%).
    The X-ray acceptance rate is also lower for the CMOS sensor.
    This difference is likely related to the detector structure of the CMOS sensor.
    In addition to the thin depletion layer, CMOS sensors include regions that are not fully depleted, where the electric field is weaker.
    As a result, charge clouds generated by incident X-rays tend to spread more significantly before charge collection. 
    Furthermore, the smaller pixel size increases the probability that an X-ray event is distributed over multiple pixels. 
    Consequently, X-ray events are more likely to be distributed over multiple pixels.
    Such broadened charge distributions can resemble those produced by charged particles in spatial extent and signal distribution, making them more difficult to distinguish and more likely to be misclassified.
    Although the radiation source and experimental conditions differ from those in the present study, this comparison further supports the notion that detector thickness and sensitive volume significantly influence achievable discrimination performance.

    In addition to the detector-dependent effects discussed above, the improved rejection performance achieved with the pnCCD-based system may also be beneficial for future scientific observations.
    Although EAGLE is primarily designed to detect and localize GRBs in the 0.4 -- 4~keV soft X-ray band, GRB spectra are known to extend over a broad energy range.
    At higher energies, where the focusing efficiency of the Lobster Eye Optics decreases, a larger fraction of detected photons may correspond to non-focused components, for which precise localization is limited. 
    Nevertheless, the thick-depletion pnCCD retains sensitivity to higher-energy photons, enabling spectral measurements beyond the nominal energy band.
    In such extended-band analyses, robust rejection of charged-particle events with large deposited energies becomes increasingly important, further highlighting the advantage of the CNN method.

\subsection{Limitations of the Classifier} 

    Despite these advantages, the CNN method does not eliminate all misclassifications.
    Two characteristic tendencies are observed. 
    
    First, events occurring near the image boundary are occasionally misclassified. 
    Although this behavior may partly reflect the limited number of edge events included in the training dataset, edge events are an unavoidable feature of practical detector operation and therefore represent an operational challenge rather than merely a training-data limitation. 
    One possible engineering solution is to apply padding strategies when constructing the input image, for example by padding missing pixels with constant values or mirrored boundary information. 
    Future work will investigate whether such approaches can improve the classification performance for edge events.
    
    Second, some $\beta$-ray events are confined within a $3\times3$ region and exhibit relatively low surrounding pixel signals, making them difficult to distinguish from genuine X-ray events.
    Such ambiguous events constitute one of the primary sources of residual misclassification in the present study. 
    In the present analysis, event labels were assigned according to the irradiation condition, with events acquired during $^{55}$Fe irradiation treated as X-ray events and those acquired during $^{90}$Sr irradiation treated as charged-particle events after applying noise-rejection and high-pulse-height selection criteria. 
    Although contamination from unrelated background events is expected to be small, some degree of label ambiguity is unavoidable. 
    Similar ambiguities are also expected in actual space observations and therefore represent an intrinsic limitation of morphology-based classification methods. 
    Future studies should evaluate the performance of the classifier using more realistic mixed-event datasets, including simulated space-background events and overlapping X-ray/charged-particle event populations, in order to assess the impact of label ambiguity and event cross-contamination on the classification accuracy.

    The CNN was trained using approximately balanced numbers of X-ray and charged-particle events in order to prevent the classifier from being biased toward the majority class and to allow it to learn representative features of both event types. 
    In the actual orbital environment, however, the event population may be dominated by charged-particle backgrounds. 
    In this case, the false-positive rate becomes particularly important, because even a small fraction of charged-particle events misclassified as X-rays can contribute to the residual background. 
    In the present validation dataset, 59 out of 1893 $\beta$-ray events were misclassified as X-ray events, corresponding to the measured misclassification rate of 3.1\%.
    Future studies using simulated orbital event populations will be required to evaluate the impact of realistic class imbalance on onboard event-selection performance.

\subsection{Limitations of the Experimental Validation} 

    The present evaluation is limited to energies above approximately 1.4~keV because of the readout noise of the current system. 
    At lower deposited energies, the number of generated charge carriers decreases, resulting in a reduced signal-to-noise ratio and less pronounced charge-distribution features. 
    Consequently, the morphological information used by the CNN for event discrimination may become less distinct. 
    In this regime, both X-ray and charged-particle events are expected to exhibit simpler charge distributions, potentially reducing the advantage of the CNN relative to conventional grade methods. 
    Therefore, although the CNN demonstrated superior performance above approximately 1.4~keV, its discrimination performance in the 0.4 -- 1.4~keV band cannot be directly inferred from the present results and remains to be quantitatively evaluated. 
    Future measurements using an improved low-noise readout system will be required to assess the classifier performance over the full mission energy range.

    In addition, the CNN model was trained using X-ray events obtained from a $^{55}$Fe source, corresponding primarily to photon energies near 6~keV. 
    Because the CNN input contains both spatial and amplitude information, the classifier may implicitly learn energy-dependent features of the charge distribution. 
    At lower X-ray energies, the charge-distribution characteristics may differ from those represented in the training dataset.
    Although the present results demonstrate the effectiveness of the CNN for events near 6~keV, its applicability to the full mission energy range cannot be established from the current dataset alone. 
    Future measurements using lower-energy X-ray sources will be required to evaluate the classifier performance in the lower-energy portion of the mission band and to retrain the model using representative low-energy X-ray events.

    The charged-particle source used in this study was a $^{90}$Sr $\beta$ source, providing electrons with energies up to 0.546 MeV. 
    To assess the relevance of this energy range to the expected orbital environment, the trapped-electron spectrum for a representative HiZ-GUNDAM orbit (600 km altitude, 98$^\circ$ inclination, solar minimum) was examined using the AE8 model in SPENVIS \cite{Heynderickx2000SPENVIS}. 
    The predicted integral flux above 40~keV is $1.09\times10^{5}$ cm$^{-2}$ s$^{-1}$, whereas the integral flux above 500~keV is $6.37\times10^{3}$ cm$^{-2}$ s$^{-1}$.
    These values indicate that lower-energy electrons constitute the dominant component of the trapped-electron environment. 
    As a result, the energy range covered by the $^{90}$Sr source overlaps with a substantial fraction of the expected orbital electron population.
    However, the discussion above considers only the energy distribution of orbital electrons.
    The charged-particle environment in orbit also includes other particle species, particularly protons.
    Because protons and electrons interact differently with the detector material, the resulting charge-distribution characteristics may not be identical, and the applicability of the present classifier to proton-induced events remains to be evaluated.
    Future studies using Geant4-based particle transport simulations and proton irradiation experiments will be performed to evaluate the classifier performance under more realistic space-radiation conditions.

    Although a comprehensive evaluation of pile-up effects has not been performed in the present study, an initial assessment was carried out using pile-up candidate events identified in the $^{55}$Fe validation dataset. 
    Pile-up candidate events were defined as events for which at least one pixel outside the central $5\times5$ event region exceeded the event threshold of 1000 ADU, indicating the possible presence of an additional nearby event.
    Using this criterion, 924 pile-up candidate events were identified. 
    Among these events, 811 (87.8\%) were classified as X-ray events by the CNN, whereas 113 (12.2\%) were rejected as charged-particle events. 
    These results indicate that the CNN accepts a substantial fraction of the pile-up candidate events present in the validation dataset.
    However, the present analysis does not distinguish between pile-up involving two X-ray events, an X-ray and a charged-particle event, or two charged-particle events.
    A more comprehensive evaluation using controlled mixed-event simulations and realistic orbital event populations will therefore be required in future work.

\section{Conclusion and Future Plans}
\label{sec:ConcandFuture}
    \subsection{Conclusion} 
    In this study, a CNN method was evaluated and compared with conventional grade methods using pnCCD datasets to address contamination of X-ray images by charged-particle events.
    Application to X-ray and $\beta$-ray datasets demonstrated an approximately four-fold reduction in the misclassification rate compared with conventional grade methods (from 10.2--11.9\% for grade methods to 3.1\% for the CNN method).
    The comparison also revealed an energy-dependent performance difference.
    Both methods showed comparable performance in the low-energy range accessible under the present experimental conditions (1.4 -- 4~keV), whereas the CNN method provided significantly improved discrimination at higher deposited energies ($>4~\rm{keV}$).
    
    The improvement originates from the ability of the CNN to utilize both the spatial structure of charge distributions and the relative signal information among pixels.
    In contrast to conventional grade methods, which rely on predefined threshold-based criteria, the CNN can capture subtle differences between compact X-ray events and charged-particle-induced signals.
    Furthermore, the comparative analysis with a thin-depletion-layer CMOS sensor indicates that the thick depletion layer of the pnCCD enhances the detectability of extended charge tracks, thereby providing more informative features for discrimination.
    This result highlights the advantage of thick-depletion detectors for CNN-based event classification.
    These results demonstrate the potential of CNN-based event classification for onboard event selection in future pnCCD-based space missions.
    
    \subsection{Future Outlook}
    To further advance the practical application of this method, several directions are planned.
    The present experimental evaluation was limited to energies above $\sim$1.4 keV due to the readout noise of the current system. 
    Future experiments using an improved readout system will extend the evaluation down to the mission energy band of 0.4--4 keV.

    In addition, cosmic-ray backgrounds in orbit consist of various particle species, including both electrons and protons, over a broad energy range. 
    Therefore, irradiation tests using high-energy protons ($\sim$100 MeV) will be conducted to assess the robustness of the classifier under realistic space-radiation conditions. 
    Complementary simulation studies using the EAGLE simulator under development by the \textit{HiZ-GUNDAM} team will also be performed. 
    By simulating electrons and protons over a wide energy range, the impact of charged-particle backgrounds on detector performance will be quantitatively evaluated.

    Long-term radiation exposure may degrade detector characteristics, such as the charge-transfer efficiency, resulting in changes to the charge-distribution morphology. 
    Such effects could alter the classification performance over the mission lifetime. 
    Future studies will investigate the impact of radiation-induced detector degradation using calibration measurements and simulated degraded detector responses.
    Periodic retraining of the classifier may provide a practical approach to maintaining stable performance throughout the mission.

    Finally, practical onboard implementation must also be demonstrated.
    The adopted CNN architecture is intentionally compact, consisting of a $5\times5$ input image, two convolutional layers, and 9,411 trainable parameters.
    The inference process requires approximately $10^4$ multiply--accumulate operations per event.
    Since the CNN is applied only to event candidates exceeding the event threshold rather than to all pixels in the image, the computational load scales with the number of detected events.
    These characteristics indicate a relatively low computational complexity for FPGA implementation.
    Nevertheless, evaluation of the processing speed, FPGA resource utilization, and power consumption remains future work.
    These studies will be conducted to assess the feasibility of real-time onboard operation.

\section*{Acknowledgment}
    This work was supported by the MEXT/JSPS KAKENHI Grant Nos. JP23H04898 (D.Y., M.A., A.D.), JP23H04891 (D.Y.), JP23H04895 (T.S.), and JP25K23398 (S.U.); the Chozen Project 2022 at Kanazawa University (D.Y., M.A., T.S.); and the Program for Forming Japan’s Peak Research Universities (J-PEAKS) at Kanazawa University (Grant No. JPJS00420230006).
    H.S. was supported by RIKEN Special Postdoctoral Researchers Program.


    \printcredits

    \bibliographystyle{unsrt}

    \bibliography{cas-refs.bib}

    \bio{}
    \endbio

\end{document}